\documentclass[prc,aps,groupedaddress,superscriptaddress,twocolumn,nofootinbib,showpacs,showkeys,floatfix,a4paper,10pt]{revtex4-1}

\usepackage{graphicx,colordvi,rotating}
\usepackage{multirow,color}
\usepackage{amsmath,amssymb}
\usepackage[mathscr]{euscript}
\usepackage{cancel}
\usepackage{braket}
\usepackage{cases}
\usepackage{color}
\usepackage{mathtools}
\usepackage{natbib}
\usepackage[normalem]{ulem}
\usepackage{graphicx}
\usepackage{dcolumn}
\usepackage{bm}
\usepackage{natbib}
\usepackage{enumerate}
\usepackage[dvipsnames]{xcolor}
\usepackage{subcaption}
\usepackage{amsmath}
\usepackage{wrapfig}

\usepackage[utf8]{inputenc}
\usepackage{epstopdf}

\usepackage{t1enc}
\usepackage[utf8]{inputenc}

\begin{document}


\title{Microscopic description of cluster radioactivity fission valleys along isotopic and isotonic chains.}

\author{M. Warda}
\email{michal.warda2@mail.umcs.pl}
\affiliation{Institute of Physics, Maria-Curie-Skłodowska Univeristy, Lublin, Poland}

\author{A. Zdeb}
\email{anna.zdeb@mail.umcs.pl}
\affiliation{Institute of Physics, Maria-Curie-Skłodowska Univeristy, Lublin, Poland}

\author{R. Rodr\'iguez-Guzm\'an}
\email{guzman.rodriguez@nu.edu.kz}
\affiliation{Department of Physics, Nazarbayev 
University, 53 Kabanbay Batyr Ave., Astana 010000, Kazakhstan}
\date{\today}

\begin{abstract}
Cluster radioactivity has been successfully described as a super-asymmetric fission mode within the 
microscopic self-consistent Gogny Hartree-Fock-Bogoliubov 
approximation [Phys. Rev. C 84, 044608 (2011)]. For nuclei
preserving the neutron-to-proton $N/Z$ ratio of the doubly magic $^{208}$Pb, a cluster radioactivity fission 
valley has been identified. Such a valley can also be found both in actinides and super-heavy nuclei. 
In this paper, chains of isotopes and isotones are examined to determine the limits of 
existence of the cluster radioactivity
fission mode. It is shown that the super-asymmetric valley can be found in a wide range
of the nuclear chart. Nevertheless, the valley flattens more and more when diverging from the 
isospin asymmetry of $^{208}$Pb. For neutron-deficient nuclei with $N/Z <$ 1.41, it is found that the valley diminishes before reaching the scission point, and cluster 
radioactivity can not be observed.

\end{abstract}


\maketitle


\section{Introduction}
\label{Introduction}

Most of the heavy and super-heavy nuclei are unstable,
with $\alpha$ emission and fission being the two foremost decay channels
(see, for example \cite{schunck16,guzman14,guzman2020,wardaPhysRevC.86.014322}, and references therein). In 1984, Rose and Jones \cite{Rose1984245} discovered an alternative decay channel by observing the 
emission of the $^{14}$C light nucleus from $^{223}$Ra. The experiment was preceded by the theoretical prediction by Sandulescu et al. \cite{Sandulescu1980528}. This decay mode is commonly known as {\it cluster radioactivity} or {\it cluster emission}. In the following 20 years, over 20 heavy isotopes have been identified as cluster emitters: from $^{221}$Fr to $^{242}$Cm, see review articles \cite{bonetti_guglielmetti_1999,Poenaru2010}. In this decay, the emitted cluster is a light nucleus with mass from $A=14$ to 34, and the heavy mass product is a double magic $^{208}$Pb or a neighbor nucleus. In all experimentally observed cases, cluster radioactivity is an exotic decay with a branching ratio to the dominant $\alpha$ emission from $10^{-9}$ to $10^{-16}$.

The theoretical description of cluster radioactivity may be carried out as an extension of the well-established Gamow model of  $\alpha$ emission or the Geiger-Nuttal law. Numerous attempts have been made in this direction over the years \cite{Zdeb2013, Tavares, renprc, qi, santosh, Krappe2012Theory}. 

An alternative approach treats this process as a super-asymmetric fission. 
In experimentally observed spontaneous fission, two types of fragment mass distributions 
can be distinguished. In symmetric fission, two identical 
fragments are produced. In the asymmetric type, the fragment mass distribution is characterized by two peaks, with the heavy-fragment mass around $A=140$, while the light one varies between $A=80$ and 120, depending on the mass of the mother nucleus. Both kinds of fission can be explained by a theory that analyzes the potential energy surface (PES) as a function of elongation and reflection asymmetry.  The fission valleys can be found on the PES between ground state and elongated shapes at large quadrupole moments.  If reflection symmetry is preserved at the fission path, it corresponds to a symmetric fission mode, whereas the valleys with non-zero octupole moment describe asymmetric fission.  However, it has 
been found that there is an alternative fission valley on the PES, leading to a very large mass asymmetry that may successfully describe cluster radioactivity \cite{War11,warda18,Matheson19,giuliani23}.  The presence of this valley is governed by the magic structure of the 
$^{208}$Pb pre-fragment, which becomes evident well before scission.

The super-asymmetric fission valley is found in actinides, where cluster radioactivity has been observed.  Both fragment mass asymmetry and decay half-lives have been reproduced reasonably well. 
It has  been shown that the barrier in this fission mode in actinides is very high (around 25 MeV), and  
that the path goes directly from the ground state to octupole-deformed configurations \cite{War11}.

The super-asymmetric path has also been found in super-heavy 
nuclei in the self-consistent calculations \cite{warda18,Matheson19} and in the 
macroscopic-microscopic approaches \cite{Ishizuka2020,Kostryukov_2021}.
In this case, the key constraint for the fissioning nucleus is that the neutron-to-proton ratio 
$N/Z=126/82\approx1.537$ of the expected heavy mass residue $^{208}$Pb shall be 
preserved roughly. 
The above condition arises from its manifestation in the experimental data of the isotopic distributions of fission fragments in actinides \cite{ramos18,ramos19}. This behavior reflects the dominant role of the symmetry energy in shaping the distribution of nuclear matter throughout the bulk of a nucleus during fission.
It turned out, upon analysis, that super-heavy nuclei around $^{284}$Cn should decay with a dominant cluster radioactivity mode \cite{warda18}. 
Experimentally, it has been found that $^{284}$Cn decays via 
fission \cite{dullmann2010,oganessian2011}. Moreover, theoretical analysis indicates that this super-asymmetric fission channel is a dominant decay mode \cite{warda18}. However, present techniques do not allow for measurements of the fission fragment mass distributions in this region.

\begin{figure}
\includegraphics[width=0.99\columnwidth,angle=0]{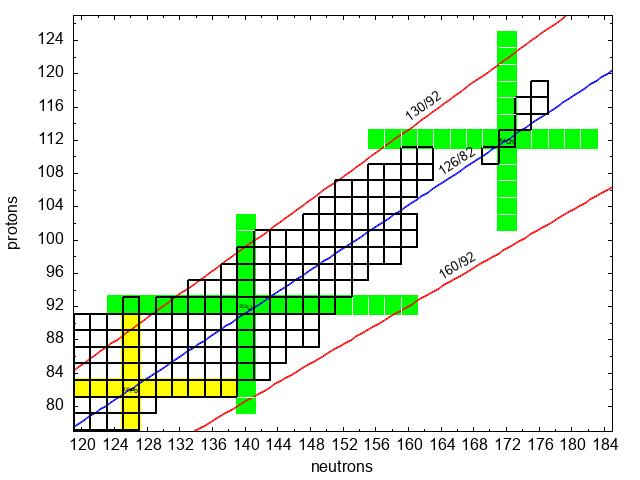}
\caption{Fragment of the nuclear chart. Black squares represent known even-even isotopes. Isotopes
 and isotones considered in this paper are marked in green. The blue line corresponds to the 
 characteristic $N/Z =$126/82 ratio for $^{208}$Pb. The red lines correspond to the 
ratios 130/92 and 160/92 for $^{222}$U and $^{252}$U.}
\label{chart}
\end{figure}

The primary objective of this paper is to examine the hypothesis that the super-asymmetric fission valley can be found only when the neutron-to-proton N/Z ratio of the parent nucleus reproduces the value of $^{208}$Pb. The existence of this valley governs the possibility of decay via the cluster radioactivity channel. To this end, we consider two isotopes in which cluster emission has been predicted and preserve the $N/Z$ ratio of $^{208}$Pb. The first one is $^{232}$U, a typical representative of the actinide region. 
Among the super-heavy nuclei we have selected, as mentioned earlier, the 
intriguing $^{284}$Cn. Building on these benchmark cases, we further investigate chains of even-even isotopes and isotones. Specifically, we consider uranium isotopes ($Z=92$) from $^{216}$U to $^{252}$U and $N=140$ isotones from $^{220}$Hg to $^{242}$No. In the super-heavy region, we study copernicium isotopes ($Z=112$) from $^{268}$Cn to $^{294}$Cn as well as $N=172$ isotones from $^{274}$No to $^{296}$124. 
A schematic representation of these nuclei in the chart of nuclides is shown in Fig.~\ref{chart}. These chains provide a framework for tracing the evolution of super-asymmetric fission valleys with increasing neutron and proton numbers and for assessing the likelihood of cluster radioactivity across both the actinide and super-heavy regions.

\section{Theory}
\label{Theory}

In this study, we have resorted to the constrained Hartree-Fock-Bogoliubov (HFB)
approximation \cite{ring1980}, with the D1S parametrization \cite{berger84} of
the Gogny energy density functional (EDF).   The
parametrization D1S has already been shown to provide a reasonable 
description of several properties, both at the mean-field level and 
beyond, all over the nuclear chart \cite{Robledo_2019}.
Here, we will briefly outline key aspects of the constrained Gogny-HFB formalism. 
For a more detailed account, the reader is referred to Ref.~\cite{Robledo_2019}.

For each of the studied nuclei, the corresponding PES 
has been obtained  with
the help of constraints on the mean value of the axially symmetric 
quadrupole $\hat{Q}_{20}$ and octupole $\hat{Q}_{30}$  
operators \cite{guzman2020,guzman12,Robledo_2012} applied to the nuclear matter distribution.  
The multipole moments are defined in the traditional way in terms of the associated Legendre polynomials $P_{lm}$ 
\begin{equation}
Q_{lm}= r^l P_{lm}[\cos (\theta)] \;.
\end{equation}
Aside from the
usual HFB constraints, a constraint on the 
operator $\hat{Q}_{10}$ is used to prevent
spurious effects due to the center of mass motion
\cite{guzman12,Robledo_2012}. Note that parity 
is allowed to be broken at any
stage of the calculation. For the solution of the (constrained) Gogny-HFB equations, an
approximate second-order gradient method 
\cite{robledo11} has been
employed. In the calculations, the Coulomb exchange term
has been considered in the Slater approximation, while the spin-orbit contribution
 to the pairing field has
been neglected. The two-body kinetic energy 
as well as the (beyond-mean-field)
rotational energy correction
\cite{Robledo_2019} have been included in the presented
results.

The HFB quasiparticle operators 
\cite{ring1980}
are expanded
in an axially symmetric (deformed) harmonic oscillator
(HO) basis. The size of the basis depends on the parameters
$N_{0}$ and $q$ \cite{guzman_2024}, and the 
basis quantum numbers are restricted by the
condition
\begin{equation}
2n_{\perp}+ |m| + \frac{1}{q}n_{z}  \le  N_{0}
\end{equation}
with $N_{0} = 15$ and $q = 1.5$. Other parameters
of the HO basis are the oscillator lengths
$b_{\perp}$ and $b_{z}$. We have optimized those oscillator lengths 
to minimize the HFB energy for each configuration 
in the PES of each isotope separately.

In the next Section, we discuss and plot the energy $E$ -- the HFB energy with the mentioned corrections relative to the ground-state energy, for each isotope, focusing on its deformation dependence rather than absolute values.


\begin{figure}
\includegraphics[width=0.90\columnwidth,angle=0]{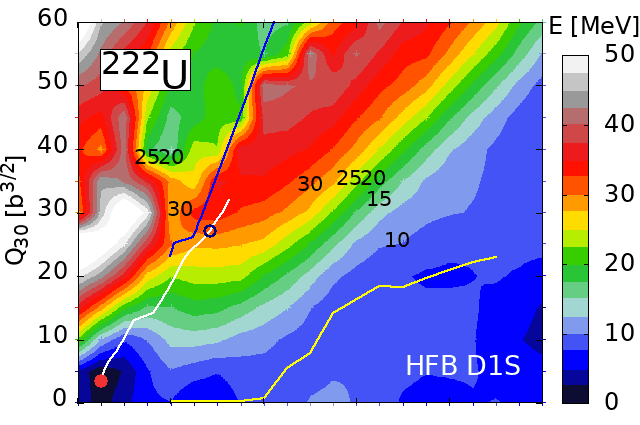}\vskip-0mm
\includegraphics[width=0.90\columnwidth,angle=0]{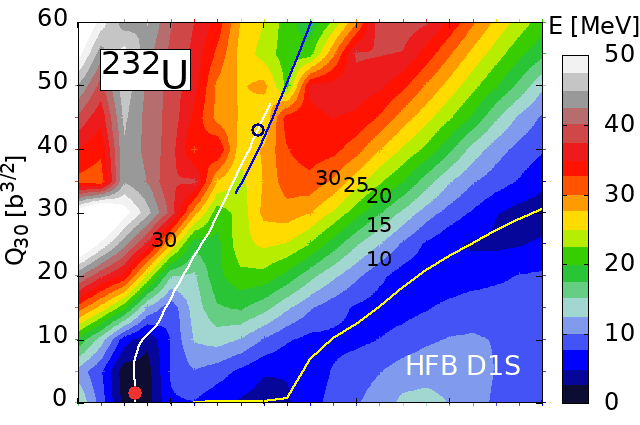}\vskip-0mm
\includegraphics[width=0.90\columnwidth,angle=0]{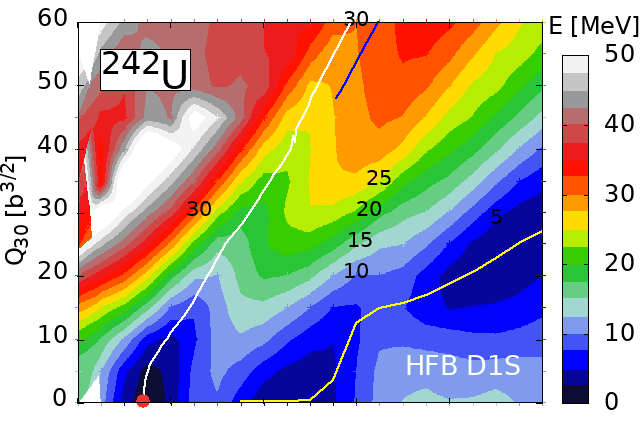}\vskip-0mm
\includegraphics[width=0.90\columnwidth,angle=0]{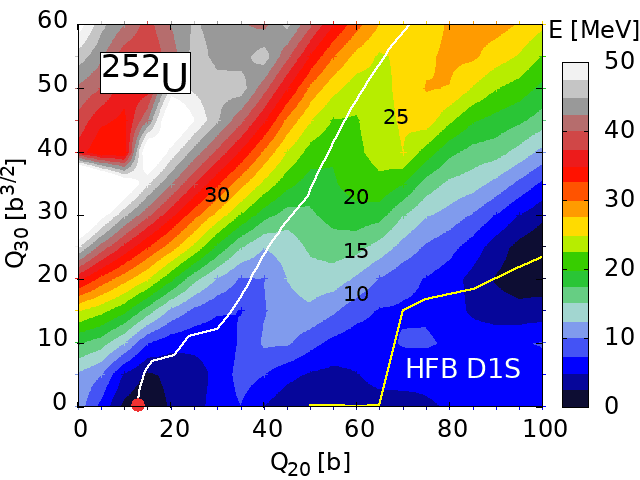}
\caption{PESs for the isotopes $^{222,232,242,252}$U. The asymmetric fission path is depicted with a yellow line. 
The super-asymmetric fission path is plotted with a white line before scission, and with a blue line after scission. The ground state is marked by a red dot, and the saddle by a black circle. }
\label{U1}
\end{figure}


\begin{figure}
\includegraphics[width=0.90\columnwidth,angle=0]{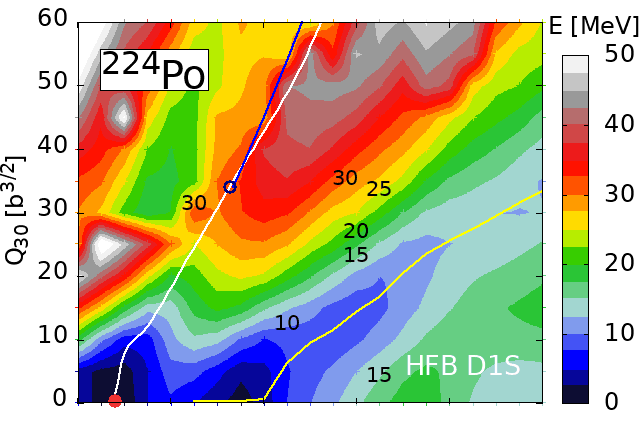}\vskip-0mm
\includegraphics[width=0.90\columnwidth,angle=0]{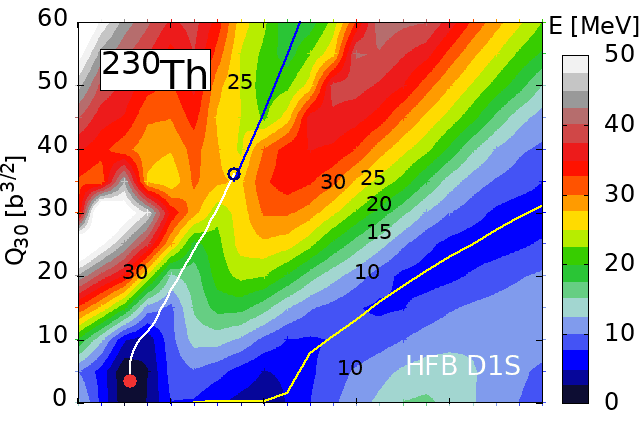}\vskip-0mm
\includegraphics[width=0.90\columnwidth,angle=0]{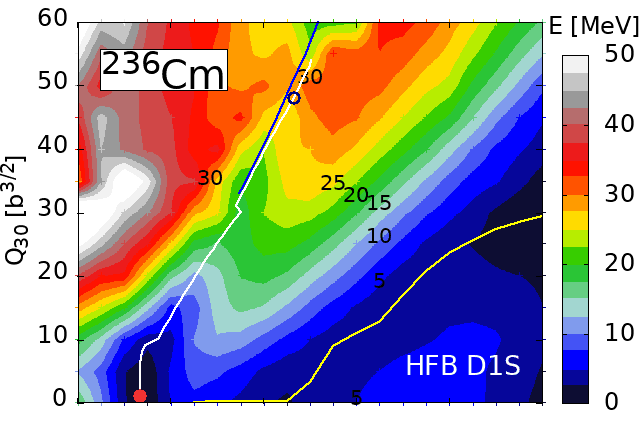}\vskip-0mm
\includegraphics[width=0.90\columnwidth,angle=0]{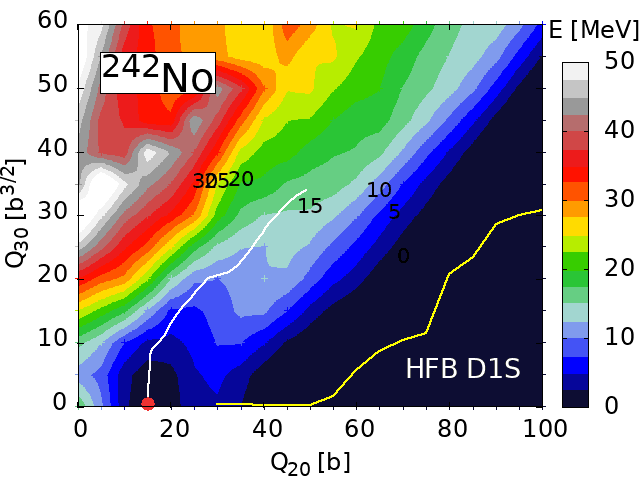}
\caption{The same as in Fig.~\ref{U1}, but for 
the $N=140$ isotones $^{224}$Po, $^{230}$Th, $^{236}$Cm and $^{242}$No.}
\label{140_1}
\end{figure}


\begin{figure}
\includegraphics[width=0.88\columnwidth,angle=0]{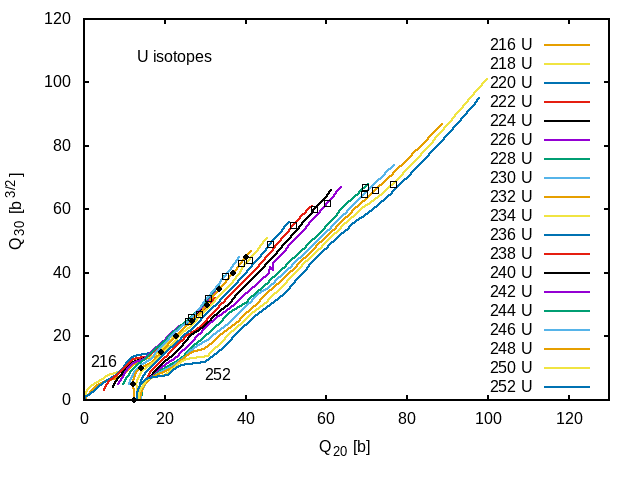}
\includegraphics[width=0.88\columnwidth,angle=0]{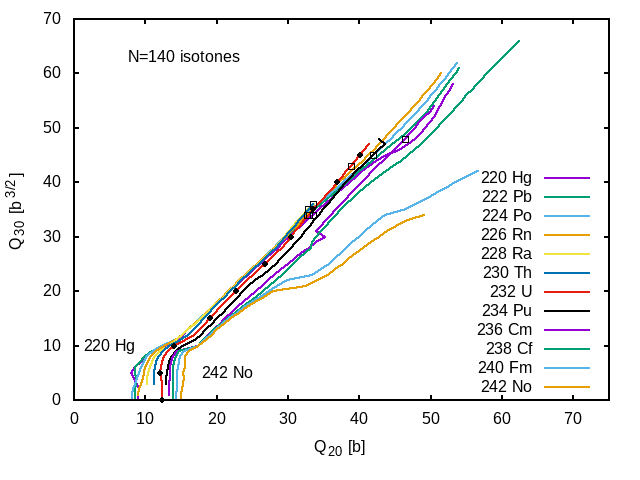}

\caption{Cluster fission paths, in the $(Q_{20}, Q_{30})$-coordinates, obtained for U
isotopes (top panel) and $N=140$ isotones (bottom panel). The path corresponding to
 $^{232}$U is highlighted by dots. Open squares mark the saddle points.  }
\label{q2q3}
\end{figure}


\begin{figure}
\includegraphics[width=0.50\columnwidth,angle=0]{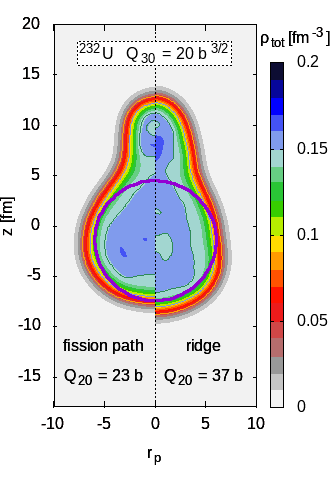}

\caption{The nuclear matter distribution of $^{232}$U at the super-asymmetric fission path $Q_{20}=23 \mathrm{b}$, $Q_{30}= 20 \mathrm{b}^{3/2}$ (left)  and at the ridge $Q_{20}=37 \mathrm{b}$, $Q_{30}= 20 \mathrm{b}^{3/2}$ (right). A circle of radius 6~fm is added to emphasize the sphericity of the heavy fragment on the path.}
\label{twoshapes}
\end{figure}


\begin{figure*}
\includegraphics[width=0.88\columnwidth,angle=0]{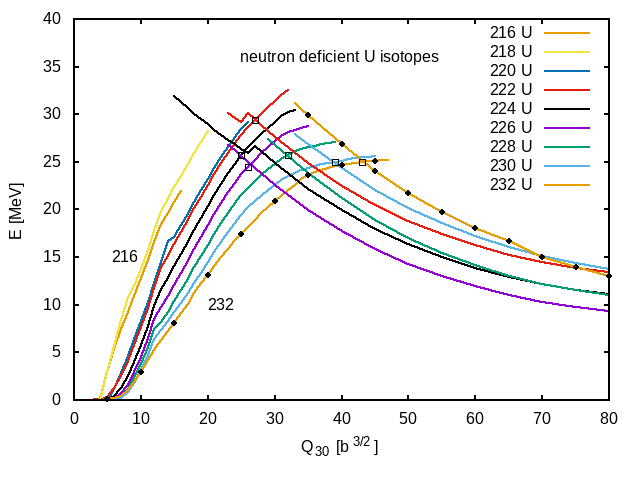}
\includegraphics[width=0.88\columnwidth,angle=0]{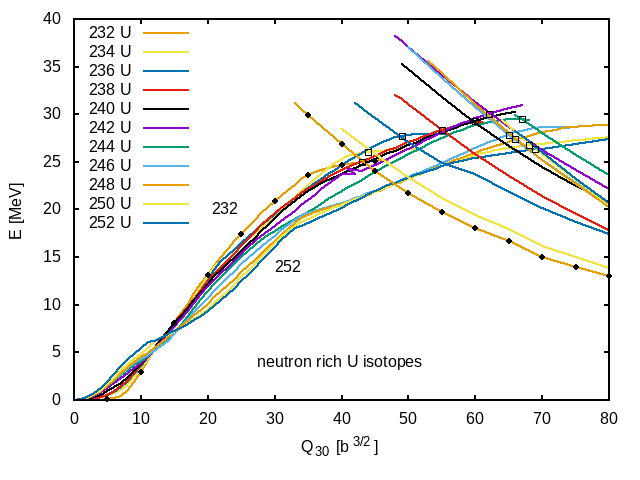}

\includegraphics[width=0.88\columnwidth,angle=0]{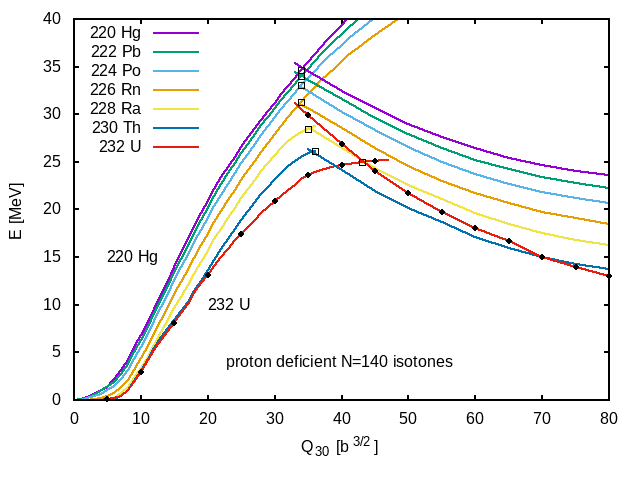}
\includegraphics[width=0.88\columnwidth,angle=0]{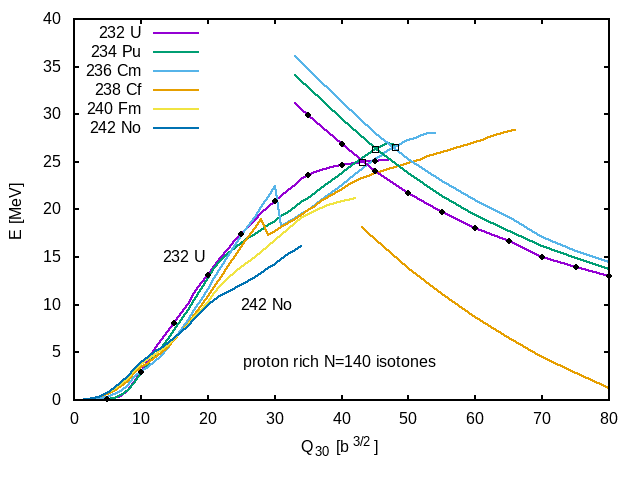}

\caption{Cluster fission barrier, as a function of the octupole moment $Q_{30}$, for neutron-deficient 
(top left panel) and neutron-rich (top right panel) U isotopes. The barriers obtained 
for proton-deficient and proton-rich $N=140$ isotones are depicted in the 
bottom left and bottom right panels, respectively. The path corresponding to  $^{232}$U is highlighted by dots in 
each panel. Open squares mark the saddle points.}
\label{q3e}
\end{figure*}

\begin{figure}
\includegraphics[width=0.485\columnwidth,angle=0]{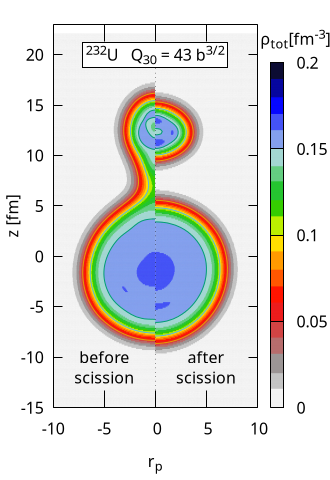}

\caption{The nuclear matter distribution of $^{232}$U at $Q_{30}=43 \;\mathrm{b}^{3/2}$ before scission (left)  and after scission (right). }
\label{scissionshapes}
\end{figure}


\begin{figure*}
\includegraphics[width=0.88\columnwidth,angle=0]{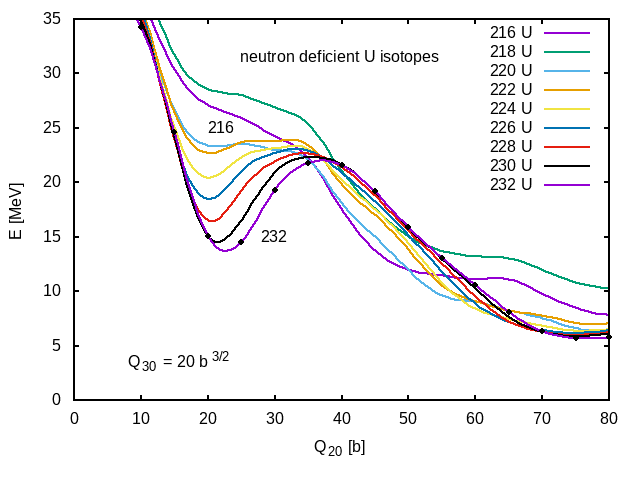}
\includegraphics[width=0.88\columnwidth,angle=0]{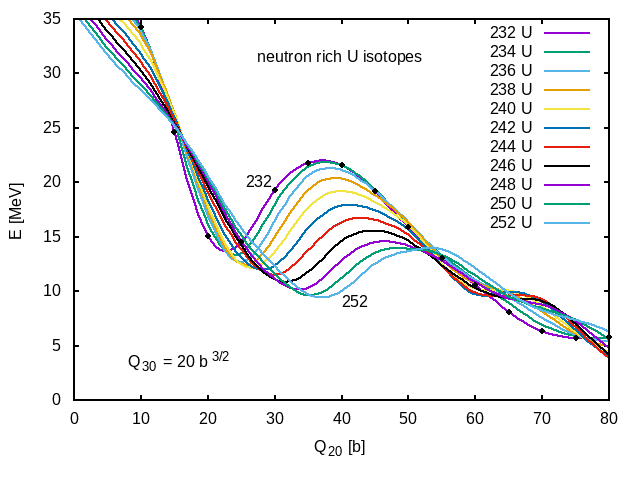}

\includegraphics[width=0.88\columnwidth,angle=0]{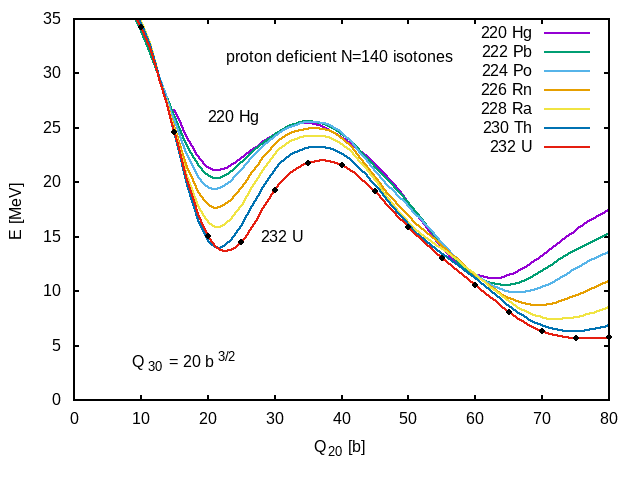}
\includegraphics[width=0.88\columnwidth,angle=0]{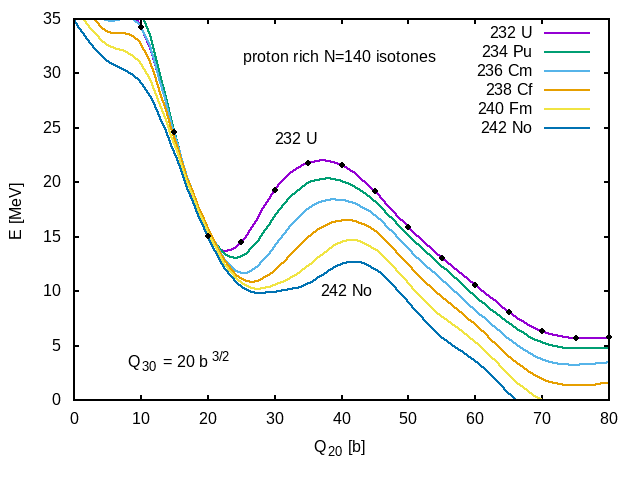}

\caption{Energy across the cluster fission valley, for a fixed value $Q_{30}=20 \;\mathrm{b}^{3/2}$ of the octupole moment, for neutron-deficient
(top left panel) and neutron-rich (top right panel) U isotopes. Results for
proton-deficient and proton-rich $N=140$ isotones are depicted in the 
bottom left and bottom right panels, respectively.  
The profile corresponding to  $^{232}$U is highlighted by dots in 
each panel.
}
\label{q30_20}
\end{figure*}

\section{Results}

In this section, we discuss the results of the calculations. First, in 
Sec.~\ref{actinides-section}, we discuss the case of actinides, while the results 
obtained for super-heavy elements are presented in 
Sec.~\ref{SH-section}.

\subsection{Actinides}
\label{actinides-section}
Let us first discuss the cluster emission valleys 
in the case of actinides. To this end, as mentioned in Sec.~\ref{Introduction}, we 
have considered the uranium isotopic chain ($^{216-252}$U)  as 
well as  $N=140$ isotones
from $^{220}$Hg up to $^{242}$No. The super-asymmetric fission valleys can be found at the 
Gogny-D1S
PESs, computed for each of the studied nuclei. As illustrative examples, the PESs obtained 
for $^{222, 232, 242, 252}$U are depicted in Fig.~\ref{U1}, and for $^{224}$Po, $^{230}$Th, $^{236}$Cm, and $^{242}$No
in Fig.~\ref{140_1}. All the PES maps can be found in the Supplemental Material \cite{supplementary}. 
One can also find there the technical details 
related to the determination of the fission paths.
As can be seen from the Figures mentioned above, the topography of the PESs has a similar structure 
for all the considered nuclei. 
 Most of the nuclei in this region exhibit a ground state quadrupole 
deformation  $Q_{20} \approx 15$~b. The only exception is the 
lightest U isotopes, for which spherical ground states are predicted. Small octupole deformations are also found in the ground states of some nuclei in this region.

The PESs reveal the asymmetric fission mode in all the considered  
actinides, i.e., a fission path that goes from the ground state towards increasing elongation with quadrupole moment. 
It is depicted with a yellow line in Figs.~\ref{U1} and \ref{140_1}. 
The inner barrier 
is axially symmetric with a height $B_{I} \le 10$ MeV. Octupole deformation plays a prominent role from the second 
minimum of the PESs, and the second barrier is asymmetric. On the other hand, the super-asymmetric fission valley can be easily distinguished. It extends from the ground state directly towards reflection-asymmetric shapes, with an almost linear relationship between the quadrupole and octupole moments. Therefore, the 
fission path at the bottom of the valley, marked by a white line, can be determined using 
a single $Q_{30}$ constraint. The cluster fission paths
in the whole region create a bunch of lines when plotted together in the 
$(Q_{20}, Q_{30})$ coordinates, as can be seen in
Fig.~\ref{q2q3}. This diagram certifies that all the considered fission valleys are  
of the same type. A slight drift towards larger quadrupole moments with increasing proton number is also apparent from the Figure.

The fission barrier in the cluster radioactivity mode in actinides has a characteristic shape \cite{War11}  consisting of two branches. In Fig. \ref{q3e}, we plotted the energy obtained for fission paths with a single constraint on the octupole moment $Q_{30}$. The first part goes from the ground state to the top of the fission barrier, at an energy of around 25 MeV. As can be seen from the left part of Fig.~\ref{twoshapes}, the molecular structure of the spherical $^{208}$Pb and the emitted cluster is visible at the early stage of the vision valley. This structure becomes increasingly visible as deformation increases. At the top of the barrier, the neck is very thin, and its rupture leads to the creation of 
two separated fragments. On the other hand, the second branch 
is associated with a system of two daughter nuclei. 
In this case, the quadrupole and octupole deformations are increased mainly by expanding the distance between the fragments without significant changes of their shape. This is accompanied by
a decrease in the Coulomb energy of the system, which governs the slope of the PES. 

Let us discuss in detail the connection between both branches of the fission barrier.
Around the point where they reach the same energy for a given octupole moment, two distinct formations appear on the PES. First is the {\it fission valley} with a molecular-like shape of a nucleus, 
and second, the {\it fusion valley } with two separated fragments. The up-going and down-going fission paths are localized at the bottoms of these valleys, respectively.
The valleys may coexist for the same constraints on the ($Q_{20}$, $Q_{30}$) plane as they are obtained in the self-consistent procedure as independent local minima of energy \cite{DUBRAY2012,zdeb2021}. 
They differ in degrees of freedom other than quadrupole and octupole deformations, e.g., in the hexadecapole moment or neck parameter \cite{berger90,War02,han21}. 
As an illustrative example in Fig. \ref{scissionshapes} we plot the density distribution of $^{232}$U at $Q_{30}=43 \;\mathrm{b}^{3/2}$ in configuration before and after scission. The heavy fragment looks similar in both cases, and the main differences lie in the presence of a neck and the deformation of the lighter fragment. The change in topology from a molecular shape with a neck to two separate fragments is clearly visible.
In consequence, the up-going and down-going branches of the barrier are not linked together at any point in the multidimensional space of deformation. 
Despite this, a continuous surface can be determined around scission when the neck parameter is applied as the third constraint \cite{War11,han21}.
More detailed analysis of the PES around the saddle is presented in the Supplemental Material \cite{supplementary}.
It was found that in the triple-constrained calculation, both valleys are coupled together, and the PES is continuous.
A barrier separating the valleys is lower than 1 MeV at the point where the branches cross 
each other for a fixed $Q_{30}$ value. 
We can make a reasonable assumption that the up-going 
and down-going branches are linked at this point, and the 
scission point is located at the saddle.

Within our microscopic Gogny-D1S HFB approach, the description of cluster decay is determined 
by the emergence of the super-asymmetric valley in the PES. Therefore, in what follows,
we will focus on the analysis of the sector of the PES from the ground state to the super-asymmetric saddle.  The super-asymmetric fission valley can be characterized by its 
length (distance from the ground state to the saddle) and its shape in the perpendicular direction. 
As shown in Figs. ~\ref{q2q3} and \ref{q3e}, the length of the fission path increases with increasing proton or neutron number in the space created by quadrupole and octupole moments.
 The scission point always remains at the same energy, around 
$25 - 30$ MeV. A slightly different trend can be noticed only in the $N=140$ isotonic chain.  The saddle point remains at the same deformation, but its energy decreases from 35 MeV to 25 MeV. This becomes apparent for 
the proton-deficient isotones, shown in the bottom left panel of Fig. \ref{q3e}. 

The key question to be addressed is the shape of the valley in a direction perpendicular to the fission path.  Is it deep or shallow, narrow or plain? 
To this end, in Fig.~\ref{q30_20} we plot the energy across the cluster decay valley for a fixed value of the octupole moment, $Q_{30}=20 \;\mathrm{b}^{3/2}$. 
In the case of the benchmark nucleus  $^{232}$U, the super-asymmetric valley is visible at $Q_{20}=23$~b. The peak of the ridge, separating from the normal asymmetric valley at 
$Q_{20}=37$~b, is 9 MeV high. 
As already mentioned above, this difference reveals the huge impact of the shell correction associated with the doubly magic nucleus $^{208}$Pb on the energy of the system. The sphericity of the heavy fragment is clear visible quite far from the scission point, see the left part of Fig. \ref{twoshapes}. One can also observe the early stage of neck formation and the molecular shape of a nucleus. Contrary, the right part of this Figure depicts the shape of the nucleus with the same octupole deformation, but on the ridge. None of the parts of a nucleus shows a shape characteristic of a magic structure. The lack of a significant shell correction gives a much larger energy than on the fission path.

Looking at  Figure \ref{q30_20}, we can analyze the evolution of the shape of the super-asymmetric valley with increasing proton and/or neutron numbers at the fixed octupole moment.
The position of the minimum in the super-asymmetric valley drifts slowly towards larger values of the quadrupole moment. The fastest drift is observed in the neutron-rich uranium isotopes. The same 
applies to the position of the ridge in this valley. This is due to the changes in the geometry of the nucleus 
with increasing mass and the Coulomb repulsion between pre-fragments. From the top left panel of 
Fig.~\ref{q30_20}, one realizes that reducing the number of neutrons 
along the uranium chain leads to a rapid increase in the energy of the path. 
This signals the vanishing of the shell structure of $^{208}$Pb due to the dominant role of the bulk symmetry energy in neutron-deficient uranium nuclei. As a consequence, the ridge disappears 
in $^{222}$U and lighter isotopes. Cluster radioactivity becomes impossible in this region 
as no preferred fission path minimizing tunneling probability can be found. On the other hand, in the case of neutron-rich 
uranium isotopes, shown in the top right panel of Fig.~\ref{q30_20}, both the energy at 
the minimum of the valley and at the ridge decrease with increasing neutron number. The latter one goes faster, leading to the reduction of the relative height of the ridge. Nevertheless, we notice that a super-asymmetric fission valley exists in these very neutron-rich isotopes. 

For all the considered $N=140$ isotones, the quadrupole moment of the minimum and the ridge remains almost unchanged.
From the bottom left and 
bottom right panels of Fig.~\ref{q30_20}, we observe that $^{232}$U has the deepest valley in the whole set. The flattening found for 
proton-deficient isotones is driven by the fast increase of the energy at the minimum of the valley. Contrary, for proton-rich nuclei, the energy at the ridge rapidly decreases with 
increasing 
proton number. Note that in $^{220}$Hg and $^{242}$No, the height of the ridge is lower than 4 MeV
 at $Q_{30}= 20 \;\mathrm{b}^{3/2}$. As can be seen from Fig.~\ref{140_1}, in the case of 
 $^{242}$No, the ridge diminishes at slightly larger octupole moment.
 
From the results discussed in this section, we conclude that 
compared to all the neighboring isotopes and isotones, the nucleus $^{232}$U has the 
deepest super-asymmetric fission valley. This clearly indicates that its $N/Z$ ratio  (which corresponds to that of $^{208}$Pb) favors the development of a molecular shape with doubly magic structure. For neutron-deficient (proton-rich) nuclei, the super-asymmetric fission valley disappears 
at
$N/Z \approx 192/90$ ($^{222}$U or $^{242}$No) [see, the red line 
in Fig.~\ref{chart}]. This seems to be the limit where the cluster radioactivity fission path can be traced 
on the PES. On the other hand, the super-asymmetric fission valley exists in all neutron-rich (proton-deficient) 
nuclei, but it is much less pronounced than in $^{232}$U when the $N/Z$ ratio increases.

\begin{figure}[hbt!]
\includegraphics[width=0.90\columnwidth,angle=0]{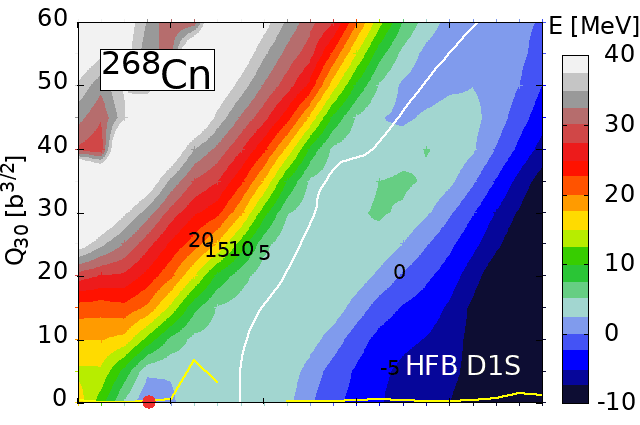}\vskip-0.cm
\includegraphics[width=0.90\columnwidth,angle=0]{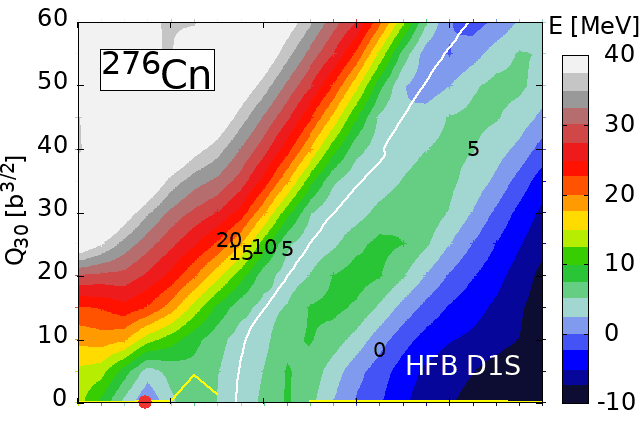}\vskip-0.cm
\includegraphics[width=0.90\columnwidth,angle=0]{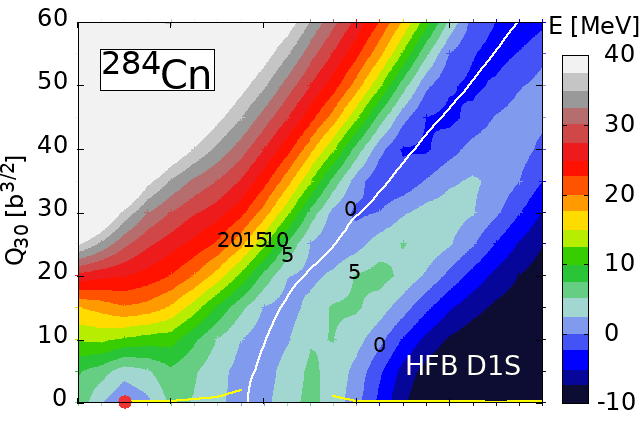}\vskip-0.cm
\includegraphics[width=0.90\columnwidth,angle=0]{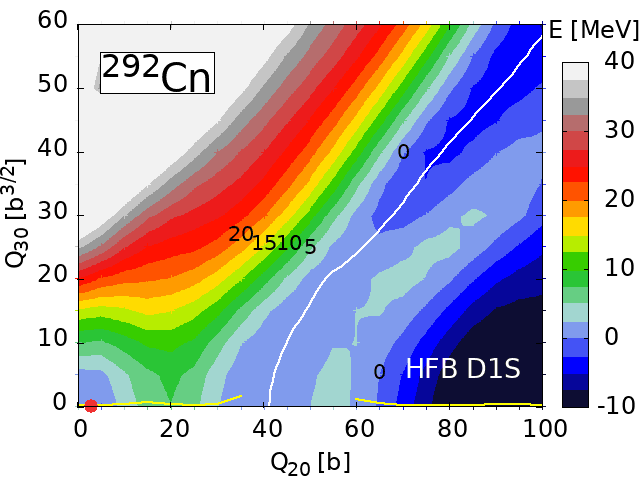}
\caption{The same as in Fig.\ref{U1} but for  the isotopes $^{268,276,284,292}$Cn. }
\label{Cn1}
\end{figure}
\begin{figure}[hbt!]
\includegraphics[width=0.90\columnwidth,angle=0]{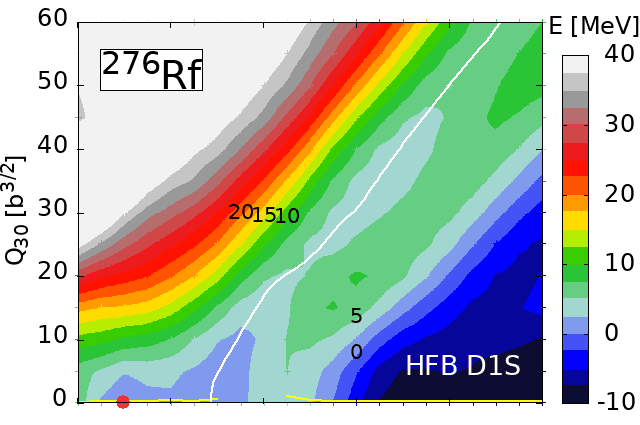}\vskip-0cm
\includegraphics[width=0.90\columnwidth,angle=0]{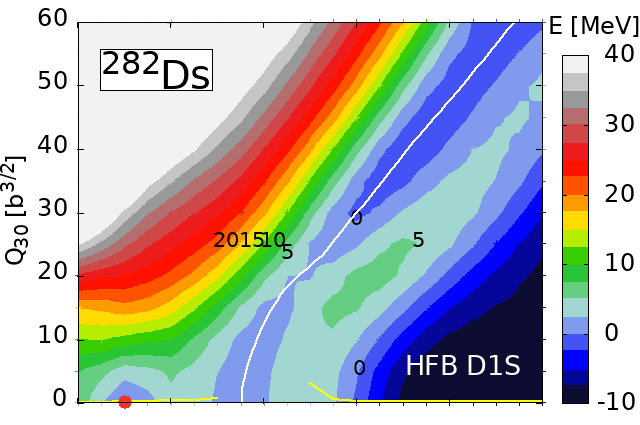}\vskip-0cm
\includegraphics[width=0.90\columnwidth,angle=0]{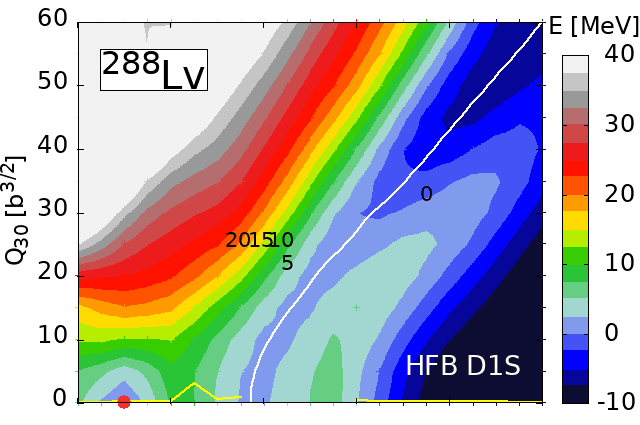}\vskip-0cm
\includegraphics[width=0.90\columnwidth,angle=0]{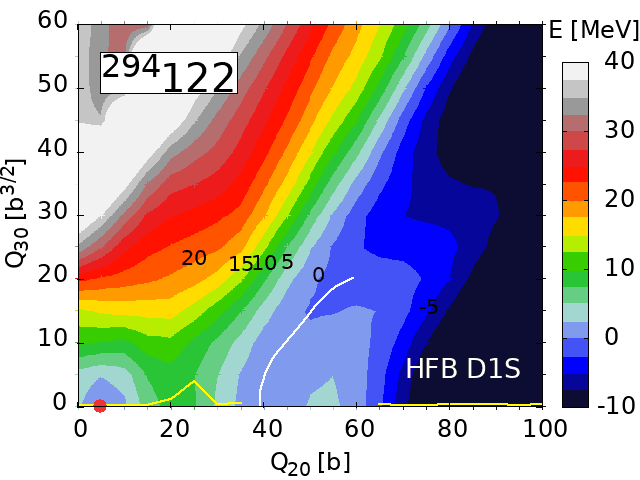}
\caption{The same as in Fig.~\ref{U1}, but for 
the $N=172$ isotones $^{276}$Rf, $^{282}$Ds, $^{288}$Lv and $^{294}$122.}
\label{172_1}
\end{figure}
\begin{figure}[hbt!]
\includegraphics[width=0.88\columnwidth,angle=0]{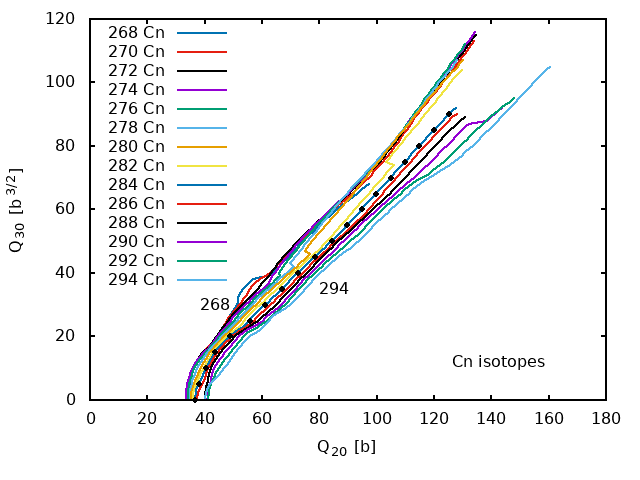}
\includegraphics[width=0.88\columnwidth,angle=0]{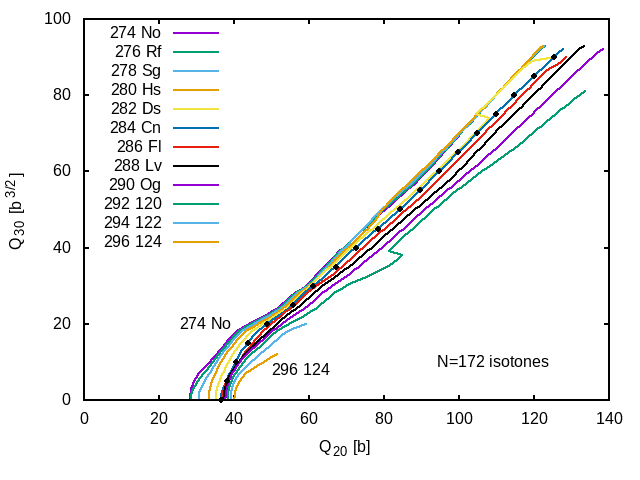}
\caption{Cluster fission paths, in the $(Q_{20}, Q_{30})$-coordinates, obtained for Cn
isotopes (top panel) and $N=172$ isotones (bottom panel). The path corresponding to
 $^{284}$Cn is highlighted by dots. The paths have been computed with a single constraint on the 
 octupole moment $Q_{30}$. }
\label{Cn-q2q3}
\end{figure}

\begin{figure}[hbt!]
\includegraphics[width=0.88\columnwidth,angle=0]{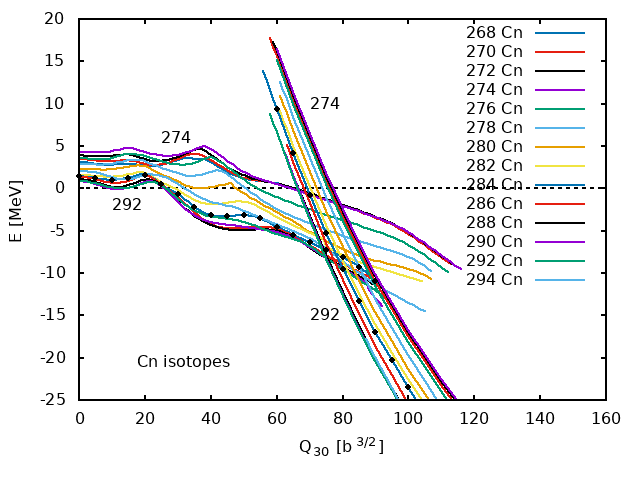}
\includegraphics[width=0.88\columnwidth,angle=0]{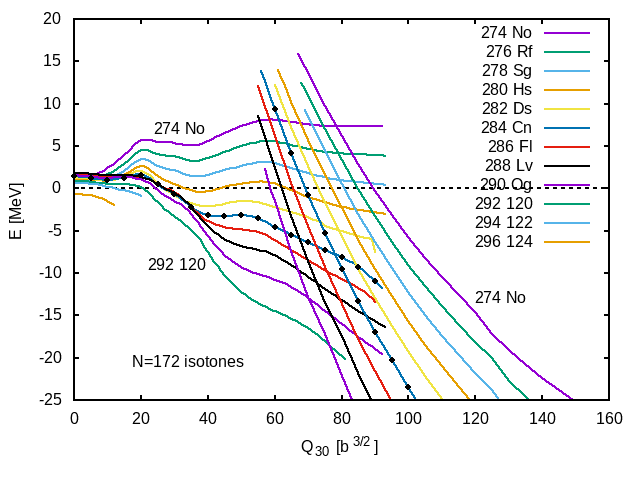}
\caption{Cluster fission paths, as functions of the octupole moment $Q_{30}$, obtained 
for Cn isotopes (top panel) and the $N=172$ isotones (bottom panel).
The path corresponding to
 $^{284}$Cn is highlighted by dots. The paths have been determined with a single constraint on 
 the octupole moment $Q_{30}$. }
\label{Cn-q3e}
\end{figure}

\begin{figure*}[hbt!]
\includegraphics[width=0.88\columnwidth,angle=0]{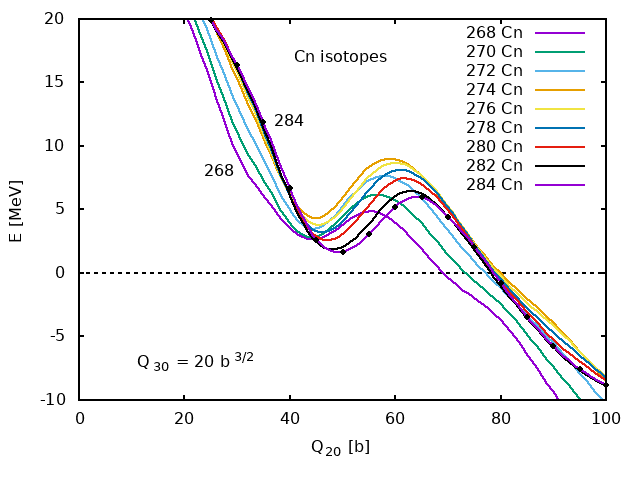}
\includegraphics[width=0.88\columnwidth,angle=0]{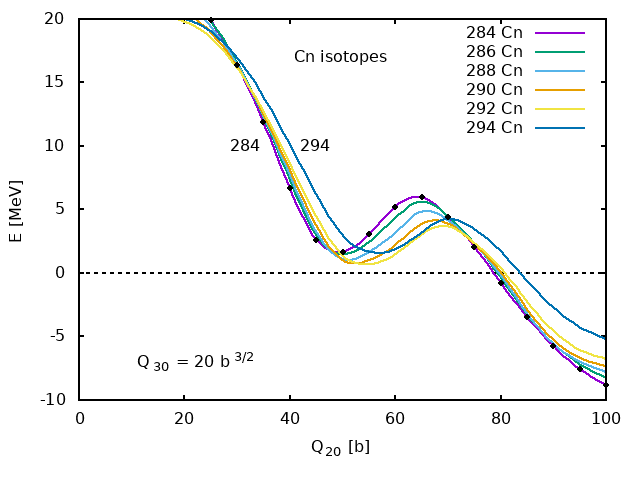}

\includegraphics[width=0.88\columnwidth,angle=0]{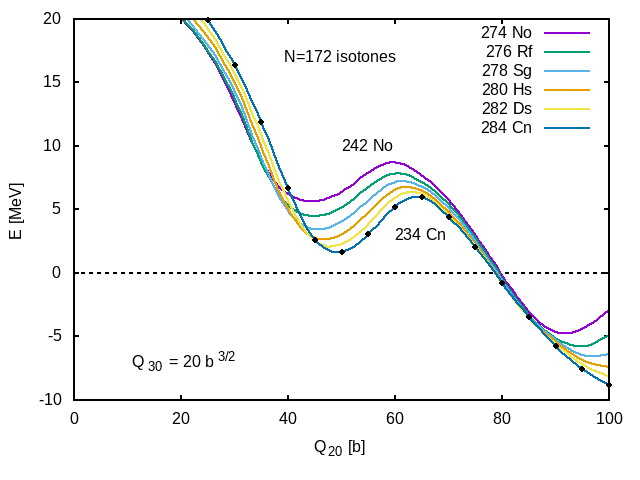}
\includegraphics[width=0.88\columnwidth,angle=0]{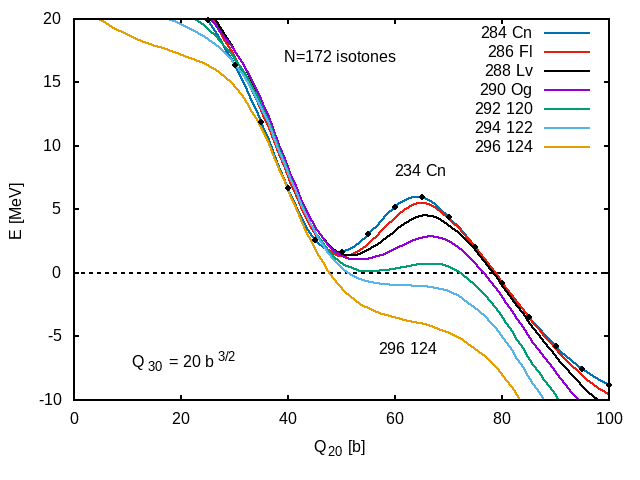}

\caption{Energy across the cluster fission valley, for a fixed value $Q_{30}=20 \;\mathrm{b}^{3/2}$ of the octupole moment,  for neutron-deficient 
(top left panel) and neutron-rich (top right panel) Cn isotopes. Results for
proton-deficient and proton-rich $N=172$ isotones are depicted in the 
bottom left and bottom right panels, respectively.  
The profile corresponding to  $^{284}$U is highlighted by dots in 
each panel.
}
\label{cn-q30_20}
\end{figure*}

\subsection{Super-heavy nuclei}
\label{SH-section}
In this section, we 
turn our attention to copernicium isotopes ($^{268-294}$Cn) as well as to $N=172$ isotones from 
$^{274}$No up to $^{296}$124. It has already been shown in Ref. \cite{warda18, Matheson19} that, in the region of super-heavy nuclei, there 
exists a fission valley in the PES that corresponds to the same kind of 
super-asymmetric fission with a doubly magic nucleus $^{208}$Pb as a heavy fragment. 
The nucleus $^{284}$Cn represents a typical example  
reproducing the $N/Z$ ratio of this residual nucleus with an open cluster emission decay channel.
Moreover, fission has been observed experimentally 
in this isotope \cite{dullmann2010,oganessian2011}, and it has been found that 
this super-asymmetric fission channel is a dominant decay mode \cite{warda18}. 
 
Despite the similarities, there are substantial differences between the cluster radioactivity valleys 
in actinides and super-heavy nuclei. The first distinction is their localization in the ($Q_{20}$, $Q_{30}$) plane. In super-heavy elements, the super-asymmetric 
path with non-zero octupole deformation does not arise from the ground state, but it starts between the first and second symmetric fission barriers. Second, the energy is much lower on the path with 
the height of the asymmetric part of the fission barrier varying between 1 MeV and 10 MeV. The third change in the energy profile is that it bends down, and the scission point is commonly below 
the ground state. This resembles the typical picture of asymmetric fission. 

For each of the studied super-heavy isotopes and/or isotones, the corresponding PES has been 
computed along the lines 
discussed in Sec .~\ref{Introduction}. The PESs obtained 
for $^{268, 276, 284, 292}$Cn as well as $^{276}$Rf, $^{282}$Ds, $^{288}$Lv and the element $^{294}$122
are depicted in Figs.~\ref{Cn1} and \ref{172_1}, as illustrative examples. All maps of the PESs are presented in the Supplemental Material \cite{supplementary}.

The light Cn isotopes are slightly elongated in their ground state. The ground state slowly shifts towards smaller values of the quadrupole moment with increasing neutron number, and the heaviest copernicium isotopes are almost spherical. 
The height of the first (axially symmetric) fission barrier gradually increases with
increasing neutron number. Looking at the PESs of the $N=172$ isotones, one can see that the ground-state position is relatively stable. The fission barrier 
increases its height with the proton number and becomes significantly wider.

The cluster fission paths, in the $(Q_{20},Q_{30})$ coordinates, obtained for the studied Cn
isotopes and $N=172$ isotones are depicted in Fig.~\ref{Cn-q2q3}. These paths 
create a tight bunch of straight lines, and
they extend to much larger quadrupole and octupole deformations than in actinides. 
A slight drift towards a larger quadrupole moment is visible with increasing mass 
number.

The energy profile of the asymmetric part of the fission barrier in the cluster radioactivity mode is shown in Fig.~\ref{Cn-q3e}. As can be seen from the top panel of the Figure, for all the studied Cn isotopes, the first part of the asymmetric path, up to $Q_{30}=20-40 \;\mathrm{b}^{3/2}$, is relatively flat in energy, while the second part goes down. The transition to the post-scission sector of the PES is possible only below the ground state energy level. For the light $N=172$ isotones, the asymmetric part of the fission barrier rises up to 9 MeV. The barrier height gradually decreases with increasing proton number. Thus, starting from $^{282}$Ds, almost the entire super-asymmetric barrier lies below the ground state energy. We have found that, in the sector where the 
neck is well developed. The fission barrier height is mainly driven by the macroscopic properties: Coulomb repulsion between the pre-fragments and the surface energy. The layout of the energy profiles resembles the results obtained in the pure liquid drop calculations \cite{ivanyuk09,ivanyuk10}.
As a result, there are substantial modifications to the proton number and modest modifications to the neutron 
number. This agrees well with the results obtained in the U region, see Fig.~\ref{q3e}.

The profiles of the cluster radioactivity  valleys, obtained for the studied isotopes and isotones 
at a fixed value $Q_{30}=20 \;\mathrm{b}^{3/2}$ of the octupole moment, are shown in 
Fig.~\ref{cn-q30_20}. From the top left and top right panels, we see that 
for isotopes from $^{274}$Cn to $^{288}$Cn, the ridge of the valley remains stable around 5 MeV above the fission path. 
For the lightest isotopes, we observe a relatively rapid reduction of the energy of the ridge, making the valley more shallow. In these nuclei, a deep valley almost converts into a plateau in the PES. 
For the heaviest isotopes, from $^{290}$Cn, the ridge also decreases its height, but it does not disappear.

It has been shown \cite{War11} that the symmetric barrier separating the beginning of the cluster radioactivity valley and the symmetric fission path can be reduced by non-axial shapes, not included in the current analysis. A small barrier obtained in the axial calculations does not guarantee blocking the evolution of the nuclear shape towards symmetric fission.

More systematic changes are visible in the profiles of the valleys obtained for 
$N=172$ isotones. In the case of the proton-deficient nuclei, bottom left panel of Fig.~\ref{cn-q30_20}, the 
height of the ridge increases with proton number due to a faster decrease of the energy at the fission path than at 
the ridge, with the highest value being reached for $^{284}$Cn. On the other hand, the 
proton-rich isotones (bottom right panel) are characterized by the fast reduction of the energy at the ridge. As a result, the fission 
path, defined as the local minimum in the calculations with a single 
constraint on the octupole 
moment, can not be found for the heaviest isotones $^{294}$122 and $^{296}$124. No cluster fission valley has been found 
for the considered value of the octupole deformation. Therefore, we conclude that the cluster emission decay channel is not set apart as a special mass asymmetry in fission. Let us
stress that the $N/Z$ ratio of $^{292}$Og corresponds precisely to the analogue limiting value in actinides.

\section{Conclusions}

In this work, we have found that super-asymmetric fission valleys exist in a wide range of heavy and super-heavy nuclei. 
It has been shown that the benchmark nuclei $^{232}$U and $^{284}$Cn have the deepest cluster radioactivity valleys among the two isotopic and two isotonic chains
considered in this work. The correlation with the
neutron-to-proton ratio of the doubly magic $^{208}$Pb becomes evident, and its shell structure 
governs the existence of the super-asymmetric fission valley. The super-asymmetric fission mode can explain the experimentally observed cluster radioactivity \cite{bonetti_guglielmetti_1999,Poenaru2010}. 

In the neutron-deficient isotopes, the valley disappears quite rapidly due to an increase in energy along the fission path (isotopic chains) or a reduction in energy at the ridge (isotonic chains). 
In the neuron-deficient part of the nuclear chart, the super-asymmetric fission valley disappears. In this region, the influence of the shell structure of the pre-fragment $^{208}$Pb is not strong enough to compete with the symmetry energy of the nuclear matter in the bulk of the fissioning nucleus.
The most neutron-deficient nuclei where the barrier can be determined are $^{222}$U, $^{236}$Cm, $^{270}$Cn, and $^{290}$Og. In these cases, the $N/Z$ ratio takes the 
values 1.41, 1.45, 1.41, and 1.46, respectively. These values are rather close to 
each other. It seems that the line $N/Z=130/92\approx1.41$  (see, Fig.~\ref{chart})
represents an upper limit for the possible existence of cluster radioactivity decay.

The cluster radioactivity valley survives in the neutron-rich region of the nuclear chart, far from the
 benchmark line defined by the $N/Z=126/82$ ratio of $^{208}$Pb. Such a valley can be identified even in nuclei that lie beyond the limit of experimental evidence of existing isotopes. The impact of the symmetry energy is not as strong as in the neutron-deficient nuclei.

In this study, we have carried out a detailed analysis of the PESs and determined the super-asymmetric fission valleys. However, it is also worth investigating in detail how the nuclear matter distribution changes along the cluster radioactivity fission paths in the isotopic and isotonic chains. This includes
modifications in the asymmetry of the proton and neutron densities as well as the formation of pre-clusters in deformed nuclei before scission. Another relevant issue is the computation of half-lives for all the nuclei considered in this study. Work along these lines is in progress and will be reported in a future publication.

\section*{Acknowledgments}

The research of M. Warda and A. Zdeb was funded by the National Science Centre, Poland, by the Grant No. 2023/49/B/ST2/01294. The research of R. Rodr\'{\i}guez-Guzm\'an was funded by Nazarbayev University 
under Faculty 
Development Competitive Research Grants Program (FDCRGP) for 2025-2027 
Grant 040225FD4712. 

The authors are grateful to Krzysztof Pomorski and Michał Kowal for careful reading of the manuscript and valuable suggestions.

\section*{Data availibilty}
The data that support the findings of this article are openly available \cite{zenodo}; embargo periods may apply.

\bibliographystyle{apsrev4-1}
%

\end{document}